\documentclass{aa}

\def\arcsspoint{\hbox to 1pt{}\rlap{\arcss}.\hbox to 2pt{}}
\def\arcsecpoint{\hbox to 1pt{}\rlap{\arcsec}.\hbox to 2pt{}}
\def\deg{\hbox{$^\circ$}}

\def\gtaprx {\lower .1ex\hbox{\rlap{\raise .6ex\hbox{\hskip .3ex
             {\ifmmode{\scriptscriptstyle >}\else
                {$\scriptscriptstyle >$}\fi}}}
                \kern -.4ex{\ifmmode{\scriptscriptstyle \sim}\else
                {$\scriptscriptstyle\sim$}\fi}}}
\def\ltaprx {\lower .1ex\hbox{\rlap{\raise .6ex\hbox{\hskip .3ex
             {\ifmmode{\scriptscriptstyle <}\else
                {$\scriptscriptstyle <$}\fi}}}
                \kern -.4ex{\ifmmode{\scriptscriptstyle \sim}\else
                {$\scriptscriptstyle\sim$}\fi}}}

\usepackage{graphics}

\begin{document}

\thesaurus{03(11.01.2, 11.10.1, 11.17.3)}

\title{Radio imaging of core-dominated high redshift quasars}

\author{Peter D. Barthel\inst{1} \and Marianne Vestergaard\inst{2,3,4}
	\and Colin J. Lonsdale\inst{5}}
\institute{Kapteyn Astronomical Institute, P.O. Box 800, NL--9700~AV
           Groningen, The Netherlands
	\and
	NBIfAFG, Copenhagen University Observatory, Juliane Maries Vej 30,
        DK--2100  Copenhagen, Denmark
	\and
	Harvard-Smithsonian Center for Astrophysics, Cambridge, MA~02138, USA
	\and
        Dept. of Astronomy, The Ohio State University, 140 W. 18th Av., 
        Columbus, OH~43210, USA (current address)
        \and
        MIT Haystack Observatory, Westford, MA~01886, USA}

\titlerunning{Core-dominated quasars}

\authorrunning{Barthel et al.}

\offprints{Peter Barthel (pdb@astro.rug.nl)}

\date{Received 16 July 1999; accepted 17 November 1999}

\maketitle

\begin{abstract}

VLA imaging at kiloparsec-scale resolution of sixteen core-dominated
radio-loud QSOs is presented.  Many objects appear to display variable
radio emission and their radio morphologies are significantly smaller
than those of steep-spectrum quasars, consistent with these objects
being observed at sight lines close to their (relativistic, $\gamma
\approx$ 4--7) jet axes.  The usefulness of the radio source orientation
indicator $R_V$, being defined as ratio of radio core and rest
frame optical V-band luminosity, is confirmed. 


\end{abstract}

\section{Introduction: classes of radio-loud AGN}

Radio-loud active galactic nuclei can to first order be classified as
having either extended (\gtaprx 15~kpc) or compact (\ltaprx 15~kpc)
radio morphologies.  The former class is characterized by steep, $\alpha
\gtaprx 0.7 \footnote{S$_{\nu} \propto \nu^{-\alpha}$}$ radio spectra,
signposting extended regions of optically thin synchrotron emission. 
Quasars or radio galaxies displaying prominent radio lobes -- on size
scales of tens to hundreds of kiloparsecs -- invariably are
characterized by steep radio spectra.  The converse, however, is not
true: compact morphologies are found in radio sources having flat as
well as steep radio spectra.  Compact steep-spectrum (CSS) radio sources
are thought to be intrinsically small objects (e.g., Fanti et al. 
\cite{fanti90}), although some CSS objects may be small due to
projection.  Ultra-compact GPS (Gigahertz Peaked Spectrum) radio sources
display inverted spectra longward and steep spectra shortward of the
peak frequency (which usually occurs around 1~GHz) -- see e.g., O'Dea
(\cite{odea98}).  It is likely that the classes of GPS, CSS, and
extended steep-spectrum sources are connected through temporal evolution
(e.g., Readhead et al.  \cite{readh96}, De Vries et al. 
\cite{devries98}, O'Dea \cite{odea98}).

Flat-spectrum radio sources invariably display a dominant compact
flat-spectrum core component, but additional weak extended emission is
usually present.  In addition, VLBI imaging usually resolves the
dominant core components into compact core-jet features, often
displaying superluminal motion.  Orr \& Browne (\cite{orr82})
originally proposed that these core-dominated sources are simply
extended, steep-spectrum objects viewed at small inclinations,
relativistic flux boosting effects being responsible for the bright
radio cores.  Imaging observations of core-dominated quasars have
yielded broad consistency with this model, as these objects usually
display weak halo and/or (curved) jet emission of small angular extent
(e.g., Kollgaard et al.  \cite{kollg90}).  Within unified models
however, the parent population of the core-dominated quasars is not
made up of lobe-dominated quasars, but of extended lobe-dominated
radio galaxies (e.g., Urry \& Padovani \cite{urpad95}).  In these
models, lobe-dominated quasars are thought to have jet inclination
angles intermediate between the radio galaxies and the core-dominated
quasars.  Having data of a large, homogeneous sample of lobe-dominated
steep-spectrum quasars at high redshift in hand (Lonsdale et al.
\cite{lbm93}, LBM93 hereafter), we here present an investigation
of this unification issue for high redshift quasars.  The uniformity
of the LBM93 data base coupled to new sensitive high resolution
observations of core-dominated objects is expected to yield useful
constraints on the relevant unification models.

\section{Sample selection and VLA observations}

\subsection{Sample selection}

In order to complement our LBM93 data base of steep-spectrum quasars, we
selected a new sample of 16 powerful radio-loud quasars with spectral
index criterion $\alpha_{408}^{5000}$ or
$\alpha_{1400}^{5000}~\ltaprx~0.5$ at redshifts $z \sim$ 2 to 3.  We
extracted these quasars from the Hewitt \& Burbidge catalog (Hewitt \&
Burbidge \cite{hb93}), using radio spectral data from the literature. 
Although our sample is not complete, having its size restricted by the
observing time allocation, we are confident that it is representative of
the population of powerful flat-spectrum radio-loud quasars at high
redshift.  Table~1 lists general information on the sample.  Most of the
radio spectral indices are in the range 0.3 -- 0.5, while three objects
with inverted spectral slope appear.  Due to radio core variability at
cm wavelengths -- common in these objects -- and non-simultaneity of the
measurements, the spectral index values will be mildly variable.  On the
basis of the information presented in Sect.~2.3 however it must be
concluded that such variability would not affect our sample selection. 

Optical QSO positions (B1950) were extracted from the literature or the
Cambridge APM facility (http://www.ast.cam.ac.uk/$\sim$apmcat/).  One
particularly inaccurate position, for \object{0504+030}, was remeasured
from a newly obtained R-band image of the QSO field, using APM star
positions.  Columns 3 and 4 of Table~1 list the optical QSO positions. 
Positional accuracies (3$\sigma$) for the optical QSOs are 1~arcsec,
except for \object{0504+030} (1{\arcsecpoint}5).  Absolute visual
magnitudes were computed from the apparent magnitudes (V\'eron-Cetty \&
V\'eron \cite{vv87}) using the cosmology\footnote{H$_0$=75, q$_0$=0 used
throughout} H$_0$=75 km~s$^{-1}$~Mpc$^{-1}$, q$_0$=0.  In addition,
these computations used optical continuum spectral index values
$\alpha=0.7$, and emission line corrections as specified in
V\'eron-Cetty \& V\'eron (\cite{vv87} -- their Fig.~2). 

On the assumption of isotropic radiation, the inferred 5~GHz radio
luminosities of these objects would be in the range P$_5 = 10^{27.4} -
10^{29.1}$ W\,Hz$^{-1}$.  Due to anisotropy of the core radiation this
assumption is most likely incorrect. As we will show below, the high
luminosities in fact imply beamed, anisotropic radiation.

\begin{table*}[!ht]
\leavevmode
\footnotesize
\caption{The flat-spectrum quasar sample}
\begin{center}
\begin{tabular}[h]{llllllr}
\hline \\[-8pt]
Quasar$^\mathrm{a}$ & Other name & RA(1950) & Dec(1950) &
 redshift & ~~$M_V$ & $\alpha_{1400}^{5000~\mathrm{b}}$ \\
\hline \\[-8pt]
\object{0106+013} & 4C\,01.02   & 01:06:04.55 & +01:19:00.3
   & 2.107 & $-$27.3 &  $0.31$  \\
\object{0123+257} & 4C\,25.05   & 01:23:57.28 & +25:43:27.8 
  & 2.356 & $-$28.2 & $-0.22$   \\ 
\object{0206+293}   & B2        & 02:06:14.97 & +29:18:34.7
   & 2.195 & $-$26.8 &  $0.34$  \\ 
\object{0226$-$038} & 4C\,$-$03.07& 02:26:22.10 & $-$03:50:58.3
   & 2.066 & $-$28.7 &  $0.25$  \\ 
\object{0317$-$023} & 4C\,$-$02.15& 03:17:56.87 & $-$02:19:26.2
   & 2.092 & $-$26.2 &  $0.47$  \\ 
\object{0458$-$020} & 4C\,$-$02.19& 04:58:41.31 & $-$02:03:33.9
   & 2.286 & $-$26.1 &  $-0.48$  \\ 
\object{0504+030}  & 4C\,03.10  & 05:04:59.24$^\mathrm{c}$ &
 +03:03:57.7$^\mathrm{c}$ & 2.453 & $-$26.9 & $0.54$  \\ 
\object{1116+128}  & 4C\,12.39  & 11:16:20.82 & +12:51:06.8
   & 2.118 & $-$26.4 &  $0.27$  \\ 
\object{1313+200}   & UT        & 13:13:58.64 & +20:02:52.5
   & 2.461 & $-$27.4 &  $0.29$  \\ 
\object{1402$-$012} & PKS       & 14:02:11.30 & $-$01:16:02.5
   & 2.522 & $-$27.8 &  $0.20$  \\ 
\object{1442+101}   & OQ\,172   & 14:42:50.58 & +10:11:12.8
   & 3.535 & $-$28.4 &  $0.54$  \\ 
\object{1542+042}  & 4C\,04.53  & 15:42:29.75 & +04:17:07.8
   & 2.182 & $-$27.6 &  $0.35$  \\ 
\object{1556$-$245} & PKS       & 15:56:41.09 & $-$24:34:11.0
   & 2.818 & $-$27.3 &  $0.40^\mathrm{d}$  \\ 
\object{1705+018}   & PKS       & 17:05:02.72 & +01:52:37.5
   & 2.577 & $-$27.3 &  $-0.27$  \\ 
\object{2048+196}   & UT      & 20:48:56.58 & +19:38:49.4
   & 2.367 & $-$27.2 &  $0.24$  \\ 
\object{2212$-$299} & PKS     & 22:12:25.09 & $-$29:59:20.0
   & 2.706 & $-$29.1 & $0.30^\mathrm{d}$  \\ 
\hline \\[8pt]
\end{tabular}
\end{center}
$^\mathrm{a}$IAU convention \\ 
$^\mathrm{b}$S$_{\nu} \propto {\nu}^{-\alpha}$ \\ 
$^\mathrm{c}$new measurement (see text) \\ 
$^\mathrm{d}\alpha_{2700}^{5000}$ \\ 
\end{table*}
\normalsize

\subsection{VLA imaging}

The flat-spectrum quasar sample was observed with the NRAO Very Large
Array, during two observing sessions: March 1990 and September 1995. 
All quasars were observed at 5~GHz (C-band), with the VLA in its high
resolution A-array configuration, yielding a typical beam size of
approximately 0.5~arcsec.  Snapshot observations combining two scans of
2--5~minutes each, at different hour angles, were obtained in order to
improve the shape of the synthesized beam. 

Secondary phase and amplitude calibrators were observed before and after
each scan.  Primary calibrator was 3C\,286, with appropriate baseline
constraints, and adopted 5~GHz flux density as provided by the VLA staff:
7.379~Jy and 7.427~Jy, at IF1 and IF2, respectively. The calibration
uncertainty is dominated by the uncertainty in the absolute flux density
of the primary calibrator, which is a few percent.  The array performed
well: judged from the calibration sources, the antenna phase and 
amplitude calibration appeared stable to within a few percent.

The radio data were of high quality, and there was no need for extensive
flagging of discrepant points.  Reduction of the data was performed
using standard NRAO AIPS image processing routines, including several
steps of self-calibration (phase only, in a few cases followed by
amplitude self-calibration).  Several successive self-calibration and
cleaning cycles generally led to a rapid convergence towards the maps
presented here.  Both full resolution maps, having synthesized beams
between 0.35 and 0.60~arcsec, and tapered maps, having beams
$\sim$1.2~arcsec were made.  In the adopted cosmology, one arcsec
corresponds to the range 8.7 -- 9.2~kpc, for the redshifts under
consideration. 

\subsection{Results and comments on individual sources}

Several sources remain unresolved at the present kpc-scale
resolution.  Images of the resolved sources are shown in Figs. 3 --
11, at the end of the paper.  Typical (1$\sigma$) image noise levels
are between 0.2 and 1~mJy/beam.  Map parameters are listed in Table~2.
All sources are individually described below.  We will frequently
compare our (epoch 1990.2 and 1995.7 -- see Table~3) 5~GHz VLA flux
densities with single dish values measured by Becker et al.
(\cite{becker91}, B91 hereafter) in epoch 1987.8, and Griffith et al.
(\cite{Griffith95}, G95 hereafter) in epoch 1990.9.

\begin{table*}[!ht]
\leavevmode
\footnotesize
\caption{Quasar contour map specifications}
\begin{center}
\begin{tabular}[h]{llrlr}
\hline \\[-8pt]
Quasar & resolving VLA beam & peak flux density & contour levels & fig. no. \\
 & (maj.axis/min.axis/p.a.) & (mJy/beam) & (mJy/beam) &   \\
\hline \\[-8pt]
0106+013   & 0.67/0.42/$-$65\deg & 1813 &
   1.0$\times(-3,3,6,12,...,1536)$  & 3 \\
0123+257   & 0.76/0.42/48\deg    & 1075 &
   1.4$\times(-3,3,6,12,...,384)$ & 4 \\
0226$-$038 & 0.45/0.37/2\deg     & 586  &
   0.5$\times(-3,3,6,12,...,768)$ & 5 \\
0317$-$023 & 0.41/0.39/$-$20\deg & 159  &
  0.25$\times(-3,3,6,12,...,384)$ & 6 \\
0458$-$020 & 0.62/0.39/5\deg     & 3543 &
  1.5$\times(-3,3,6,12,...,1536)$  & 7 \\
0504+030   & 0.46/0.39/22\deg    & 332  &
  0.25$\times(-3,3,6,12,...,768)$ & 8 \\
1116+128   & 0.54/0.40/$-$85\deg & 1673 &
  1.1$\times(-3,3,6,12,...,768)$ & 9 \\
1313+200   & 0.40/0.38/$-$51\deg & 268  &
  0.6$\times(-3,3,6,12,...,384)$  & 10 \\
1542+042   & 0.49/0.43/$-$27\deg & 457  &
  0.3$\times(-3,3,6,12,...,768)$ & 11 \\
\hline \\[8pt]
\end{tabular}
\end{center}
\end{table*}
\normalsize

\subsubsection{\object{0106+013}}

We measure a 5~GHz core flux density of 1.80~Jy and an additional flux
density of 0.15~Jy in extended emission.  Comparison with other
measurements (B91: 3.47~Jy; G95: 2.08~Jy) indicates strong
variability.  The radio structure of 0106+013 is a 4.5~arcsec double,
with a dominant core coincident with the optical QSO (Fig.~3).  Our
image is morphologically consistent with the 5~GHz VLA image of
Kollgaard et al. (\cite{kollg90}) and the lower resolution image of
Murphy et al.  (\cite{murphy93}).  The flux density contrast with the
Kollgaard et al.  data (core flux density 3.87~Jy, at observing epoch
1986.45) is nevertheless striking.

\subsubsection{\object{0123+257}}

We measure a 5~GHz integrated flux density of 1.13~Jy, of which
1.08~Jy is in an unresolved nucleus and $\sim$0.05~Jy in extended
emission, including an unresolved component, 1.0~arcsec SE of the
former (Fig.~4).  Comparison with B91, measuring 1.33~Jy, indicates
moderate 5~GHz variability.

\subsubsection{\object{0206+293}}

This (B2) quasar is unresolved, hence the radio map is not shown.  Our
5~GHz flux density of 0.28~Jy is consistent with the 1987 value (B91).

\subsubsection{\object{0226$-$038}}

We measure a 5~GHz integrated flux density of 0.65~Jy, of which
0.58~Jy is in an unresolved nucleus and 0.07~Jy in additional
components, straddling the former (Fig.~5).  The overall angular size
is 3.5~arcsec.  As measured by Wall (\cite{wall72}), Parkes single
dish data (0.55~Jy) are indicative of mild variability.

\subsubsection{\object{0317$-$023}}

We detect a 5~GHz integrated flux density of 0.26~Jy, of which 0.16~Jy
is in an unresolved nucleus and 0.10~Jy in additional components,
towards the south and in a curved small scale jet towards the NE
(Fig.~6).  The overall angular size of 0317$-$023 is 7~arcsec.  Neither
the Green Bank single dish flux density (0.28~Jy -- G95) nor the earlier 
Parkes figure (0.29~Jy -- Wall \cite{wall72}) are indicative of variability. 

\subsubsection{\object{0458$-$020}}

Well-known variable object at cm-mm wavelengths (e.g., K\"uhr et al.
\cite{kuehr81}, G95), and `marginal' EGRET $\gamma$-ray source
(Thompson et al. \cite{thomps95}).  Our integrated 5~GHz flux density
of $\sim$3.7~Jy implies a factor two of increase in comparison to
earlier data (K\"uhr et al.  \cite{kuehr81}).  A strong core (3.55~Jy
at 5~GHz) is straddled by $\sim$0.15~Jy extended emission, yielding an
overall angular size of $\sim$3.5~arcsec (Fig.~7).  A tapered map
demonstrates the reality of the faint emission NE of the nucleus.  Our
image is consistent with a MERLIN 408~MHz image (F.~Briggs, priv.
comm.; image also shown in Punsly \cite{punsl95}).

\subsubsection{\object{0504+030}}

We detect a 5~GHz integrated flux density of 0.48~Jy, of which 0.32~Jy
is in an unresolved nucleus and 0.16~Jy in a jet-hotspot structure
towards the NW as well as faint emission towards the SE (Fig.~8).  A
tapered map demonstrates the reality of this `counterjet' emission.
The overall angular size of \object{0504+030} measures 4~arcsec.  Our
flux density determination is consistent with earlier ones (B91:
0.45~Jy, G95: 0.50~Jy), indicating little or no variability.  In order
to resolve a positional discrepancy for the optical QSO, we analyzed
images of the field, taken using the 90cm ESO Dutch telescope on La
Silla.  Although the identification is secure now, we were unable to
improve on the accuracy of the astrometry, due to the small number of
suitable stars in the field of the QSO. We refer to Table~1 and the
accompanying paragraph in Sect.~2.1.

\subsubsection{\object{1116+128}}

Our measurements in comparison with literature data indicate fairly
strong core variability.  We detect 1.7~Jy in an unresolved core, and
0.1~Jy in a secondary component 2.7~arcsec towards the NW (Fig.~9).
Our integrated 5~GHz flux density of 1.8~Jy is inconsistent with
earlier values, which are in the range 1.3 -- 1.6~Jy (K\"uhr et
al. \cite{kuehr81}).  Our image is consistent with the lower
resolution image of Murphy et al.  (\cite{murphy93}), although the
latter shows some evidence for additional low surface brightness halo
emission.  O'Dea et al. (\cite{odea88}) also find evidence for diffuse
emission, besides the unequal double morphology (having 1.5~Jy in the
unresolved core).

\subsubsection{\object{1313+200}}

This object originated in the University of Texas 365~MHz survey
(Wills \& Wills \cite{ww79}).  Our integrated 5~GHz flux density of
0.29~Jy -- of which 0.27~Jy is in an unresolved component -- is not
indicative of strong variability by comparison with B91 (0.33~Jy).
Our image shows an unresolved core and faint extended emission,
0.7~arcsec NNW (Fig.~10).

\subsubsection{\object{1402$-$012}}

An unresolved source, having 5~GHz flux density 0.34~Jy.  Wall
(\cite{wall72}) reports a Parkes value of 0.81~Jy: the object must be
strongly variable.  Deconvolution gives evidence for small resolution
effects in direction N-S. The radio map is not shown.

\subsubsection{\object{1442+101}}

This is a well known GPS object, identified with the pre-1982 QSO
redshift record holder OQ\,172.  The radio source is unresolved on the
0.1~arcsec scale, hence the radio map is not shown.  VLBI observations
on the other hand have revealed a highly bent core-jet structure on the
milliarcsec scale (Udomprasert et al. \cite{udomp97}).  Our flux density
measurement of 1.22~Jy is consistent with earlier ones, e.g., K\"uhr
et al. (\cite{kuehr81}), thereby indicating little or no variability.

\subsubsection{\object{1542+042}}

We measure a 5~GHz core flux density of 0.45~Jy and an additional flux
density of 0.04~Jy in extended emission.  Most of this extended emission
originates in a secondary component 1.3~arcsec E of the nuclear
component (Fig.~11).  There is evidence for additional low surface
brightness halo emission both E and W of the nucleus, which is confirmed
by a tapered image.  Other measurements (B91: 0.41~Jy; G95: 0.54~Jy;
Shimmins et al. \cite{shimm75}: 0.47~Jy) are indicative of a small level of
variability. 

\subsubsection{\object{1556$-$245}}

We measure an unresolved 5~GHz flux density of 0.38~Jy; the radio map
is not shown.  Comparison with Parkes data (0.54~Jy -- Bolton et
al. \cite{bolt75}) indicates fairly strong variability.

\subsubsection{\object{1705+018}}

We measure an unresolved 5~GHz flux density of 0.57~Jy; the radio map
is not shown.  Comparison with earlier data (e.g., Wall \cite{wall72}:
0.58~Jy; B91: 0.42~Jy; G95: 0.46~Jy) yields evidence for moderate
levels of variability.

\subsubsection{\object{2048+196}}

We measure unresolved 5~GHz flux density of 0.11~Jy for this UT quasar,
consistent with B91; the radio map is not shown. 

\subsubsection{\object{2212$-$299}}

We measure unresolved 5~GHz flux density of 0.41~Jy, consistent with the
Parkes value (Wall et al. \cite{wall76}); the radio map is not shown.

\subsection{Source parameters}

Measured parameters for the 16 quasars, as well as inferred quantities
appear in Table~3.  The flux densities were measured in various ways.
Gaussian component fitting on the images was employed to determine the
(unresolved) core flux densities.  Flux density figures for the weak
extended emission features were determined by adding the relevant
CLEAN components in combination with examination of the short baseline
amplitudes in the $uv$-plane.  The estimated uncertainty in these
figures (3$\sigma$) is 5--10 percent.  Source intrinsic parameter
computation used the adopted cosmology.  We obviously do not list
luminosities for undetected extended emission associated with
unresolved objects.  However, taking 20~mJy as a conservative flux
density upper limit for such, probably diffuse low surface brightness
emission, the relevant luminosity upper limit is $\sim 10^{27}$
W{\thinspace}Hz$^{-1}$ at the redshifts involved, which is a
substantial value.  The penultimate column in Table~3 specifies the
radio core fraction at 5~GHz emitted frequency, log\,$R_5$.  These
figures were computed from the entries in columns 3 and 4, adopting
canonical spectral index values of 0.80 and 0 for the extended and
core emission, respectively.  The final column lists values of the
$R_V$ parameter, introduced by Wills \& Brotherton
(\cite{wibroth95}), and defined as the ratio of the core radio
luminosity at 5~GHz emitted frequency to the (K-corrected) optical
V-band luminosity as computed from the absolute visual magnitudes.  It
is easy to show that log$R_V$ = log(L$_\mathrm{core,5}$) + $M_V$/2.5
$-$ 13.7. The next section will discuss this $R_V$ parameter.

\begin{table*}[ht!]
\leavevmode
\footnotesize
\caption{Quasar properties, measured/calculated from 5~GHz VLA data}
\begin{center}
\begin{tabular}[h]{llrrlllrrr}
\hline \\[-8pt]
Quasar & epoch & flux~density & extended & 
 log(L$_\mathrm{total})^\mathrm{a}$ &
 log(L$_\mathrm{ext.})^\mathrm{b}$  &
 ang. size & lin. size &
 log\,$R_5^\mathrm{c}$ & log\,$R_V^\mathrm{d}$ \\
 & (VLA  & (mJy) & flux~density & (W/Hz) & (W/Hz) & (arcsec) & (kpc) &  &  \\
 &  obs.)&       & (mJy)   &        &        &          &       &  &  \\
\hline \\[-8pt]
\object{0106+013}   & 1990.2 & 1950 & 150 & 28.51 & 27.63 & 4.5 & 39.1 &
  $-$0.08 & 3.68 \\
\object{0123+257}   & 1990.2 & 1130 & 50  & 28.12 & 27.30 & 1.0 & 8.8  &
  $-$0.05 & 3.22 \\ 
\object{0206+293}   & 1990.2 & 280  & --$^\mathrm{e}$ & 27.73 & & & $<1$ &
  0.00    & 3.15 \\ 
\object{0226$-$038} & 1990.2 & 650  & 70  & 27.98 & 27.28 & 3.5 & 30.3 &
  $-$0.11 & 2.62 \\ 
\object{0317$-$023} & 1990.2 & 260  & 100 & 27.70 & 27.45 & 7.0 & 60.7 &
  $-$0.41 & 3.08 \\ 
\object{0458$-$020} & 1990.2 & 3700 & 150 & 28.47 & 27.74 & 3.5 & 30.8 &
  $-$0.05 & 4.55 \\ 
\object{0504+030}   & 1990.2 & 480  & 160 & 28.19 & 27.85 & 4.0 & 35.5 &
  $-$0.37 & 3.27 \\ 
\object{1116+128}   & 1990.2 & 1800 & 100 & 28.46 & 27.46 & 2.7 & 23.5 &
  $-$0.06 & 4.02 \\ 
\object{1313+200}   & 1995.7 & 290  & 20  & 27.84 & 26.96 & 0.6 & 5.3  &
  $-$0.08 & 2.99 \\ 
\object{1402$-$012} & 1995.7 & 340  & --$^\mathrm{e}$ & 27.89 & & & $<1$ &
  0.00    & 2.97 \\ 
\object{1442+101}   & 1990.2 & 1220 & --$^\mathrm{e}$ & 29.05 & & & $<1$ &
  0.00    & 3.42 \\ 
\object{1542+042}   & 1990.2 & 490  & 40  & 27.97 & 27.10 & 1.3 & 11.4 &
  $-$0.09 & 3.00 \\ 
\object{1556$-$245} & 1995.7 & 380  & --$^\mathrm{e}$ & 28.18 & & & $<1$ &
  0.00    & 3.31 \\ 
\object{1705+018}   & 1995.7 & 570  & --$^\mathrm{e}$ & 27.88 & & & $<1$ &
  0.00    & 3.43 \\ 
\object{2048+196}   & 1995.7 & 110  & --$^\mathrm{e}$ & 27.35 & & & $<1$ &
  0.00    & 2.63 \\ 
\object{2212$-$299} & 1995.7 & 410  & --$^\mathrm{e}$ & 28.11 & & & $<1$ &
  0.00    & 2.61 \\ 
\hline \\[8pt]
\end{tabular}
\end{center}
$^\mathrm{a}$K-corrected using spectral index for integrated emission \\ 
$^\mathrm{b}$K-corrected adopting spectral index 0.80 for extended emission \\ 
$^\mathrm{c}$logarithm of fractional core flux density at 5\,GHz 
 emitted frequency, computed adopting spectral indices 0.80 and 0 for 
 the extended and core emission respectively \\ 
$^\mathrm{d}$logarithm of ratio of radio core luminosity at 5\,GHz
 emitted frequency to (K-corrected) optical V-band luminosity \\
$^\mathrm{e}$diffuse extended flux density up to about 20~mJy cannot be
 excluded -- see text, Sect.2.4 \\
\label{parmtab}
\end{table*}
\normalsize

\section{Discussion}

Out of the sample of 16 core-dominated objects, nine objects display
extended emission, generally morphologically asymmetric with respect
to the dominant nuclear emission.  This is entirely in agreement with
earlier studies (e.g., O'Dea et al.  \cite{odea88}, Perley et al.
\cite{perley82}), and in marked contrast to the symmetric double-lobed
morphologies commonly seen in extended steep-spectrum radio-loud QSOs
(e.g., LBM93).  The linear sizes of the core-dominated sources are
significantly smaller: we measure a median value 7~kpc and maximum
60~kpc, which compares to 79 and 280~kpc for a subsample of 35
steep-spectrum ($\alpha \gtaprx 0.5$) LBM93 quasars, defined as having
log\,L$_{\rm ext.} \geq 27.0$. The latter figure is dictated by the
detection limit of extended emission associated with the
core-dominated quasars.  Considering the core fractions, expressed in
the $R_5$-parameter, the samples obviously differ: we find median
log\,$R_5$ values of $-0.055$ and $-1.28$ for the flat- and
steep-spectrum sample respectively.  While the former range from 0.0
to $-0.41$, the latter range from $-0.05$ to $-2.87$ (core fraction
0.13\%).  The radio luminosities of the extended emission differ much
less markedly: median values are log\,L$_{\rm ext.}  = 27.2$ and 27.7
for the flat- and steep-spectrum sample respectively.  The former
value is still two orders of magnitude in excess of the Fanaroff \&
Riley (\cite{fr74}) break luminosity.  This confirms earlier findings
by O'Dea et al. (\cite{odea88}), Murphy et al. (\cite{murphy93}), and
Kollgaard et al. (\cite{kollg90}): on the basis of their extended
emission, many core-dominated, flat-spectrum radio sources need to be
classified as Fanaroff and Riley class II sources, provided the
extended emission is isotropic.  As for the seven unresolved sources
from our sample, we cannot as yet exclude the FR\,II classification.

Hence the major differences between the present flat-spectrum sample
and the comparison sample of steep-spectrum quasars are the relative
strengths of their radio cores and their morphological (a)symmetry and
overall size.  Variability data (e.g., factors of two on time scales
of years) lead to brightness temperatures in excess of $10^{13}$K,
pointing towards beamed core emission (e.g., Kellermann \&
Pauliny-Toth \cite{kpt81}).  To this should be added the possible
gamma-ray detection of 0458+020, most likely also signaling beamed
radiation (e.g., Barthel et al.  \cite{pdb95}).  Therefore, rather
than classifying the objects as intrinsically small sources, we prefer
classification as normal extended double sources displaying beamed
core emission and lobe foreshortening due to close alignment with the
sight line of their radio axes.

\begin{figure*}
 \resizebox{12cm}{!}{\includegraphics{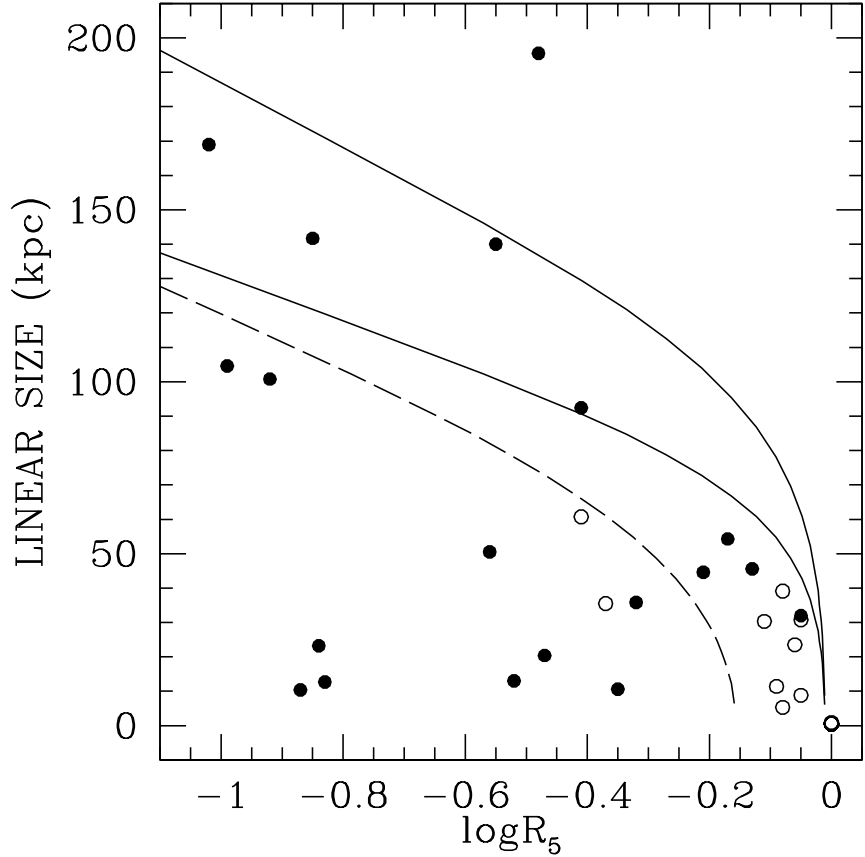}}
 \hfill
 \parbox[b]{55mm}{
  \caption{Linear size vs. core fraction at 5~GHz for the combined
   sample of core- and lobe-dominated quasars, having
   log\,L$_\mathrm{ext.} \geq 27.0$, and $-1 \leq {\rm log}\,R_5 \leq 0$.
   Open circles correspond to the observed flat spectrum quasars 
   (Tables 1 and 3).
   The two solid lines describe the behaviour of a 500~kpc (350~kpc in
   lower curve) triple radio source with relativistic ($\gamma=7.1$ or
   $\beta=.99$) nuclear jets giving rise to beamed core emission and
   foreshortened sizes at small inclination angles.  The dashed curve
   describes the behaviour of a 350~kpc radio source with $\gamma=4.1$
   jets.  The unbeamed core fraction was fixed by dictating
   log\,$R_5(\theta=45\deg) = -2.5$.  See text for details on data
   points and models.}
  \label{linR5}}
\end{figure*}

Following Kapahi \& Saikia (\cite{ks82}) we next investigate whether
the linear sizes and core prominences support this view.
Fig.~\ref{linR5} displays the linear size -- log\,$R_5$ behaviour
for the present 16 sources, supplemented with 19 objects having
log\,$R_5$ in the range $-1.0$ to 0.0 -- the interesting range for
comparison -- from the subsample of 35 LBM93 steep-spectrum quasars
described above.  The steep-spectrum objects are: \object{0225$-$014},
\object{0238+100}, \object{0352+123}, \object{0445+097},
\object{0751+298}, \object{0758+120}, \object{0805+046},
\object{0941+261}, \object{1023+067}, \object{1055+499},
\object{1354+258}, \object{1402+044}, \object{1540+180},
\object{1554$-$203}, \object{1607+183}, \object{1701+379},
\object{1726+344}, \object{2223+210}, and \object{2338+042}.  It
should be noted that for these objects the $R_5$-values were evaluated
using their measured spectral indices.  It is clearly seen that
high-$R_5$ objects commonly display extended emission.  However, this
emission does not exceed a few tens of kpc projected linear size.
Low-$R_5$ objects display a considerably wider linear size range.  The
model lines describe the predicted behaviour of a triple (core plus
two lobes) radio source with opposite relativistic nuclear jets, and
intrinsic size $L$.  With decreasing inclination angle, the effects of
foreshortening and beamed core emission will increase, hence
increasing log\,$R_5$ with decreasing linear size in the model
behaviour.

Defining $\theta$ as the angle between the radio source (jet) axis and
the line of sight, a radio source having intrinsic linear size L, 5~GHz
core fraction $F_c$ and opposite jets with bulk flow speed $\beta=v/c$
will display a projected length

$$ L(\theta) = L\,\sin(\theta)$$

as well as a core fraction (in the emitted frame)

$$ R(\theta) = [2\,(1/{F_c} - 1)/B(\theta) + 1]^{-1} $$

where 

$$B(\theta) = (1-\beta\cos\theta)^{-2-\alpha} + 
(1+\beta\cos\theta)^{-2-\alpha}$$

We adopt $\alpha=0.8$ for the jet spectral index, and jet flow speed
$\beta=0.99$ (i.e., $\gamma=7.1$) in evaluating the model, shown in
Fig.~\ref{linR5}, for intrinsic source sizes $L$ of 350 and 500~kpc,
respectively. Within current unification schemes (e.g., Barthel
\cite{pdb89}) quasars are observed within jet inclination angles
$\approx45\deg$.  LBM93 data show the largest quasars to have
log\,$R_5$ values $\sim -2.5$.  On the basis of this finding the
unbeamed core fraction $F_c \equiv R(\theta=90\deg)$ was fixed by
dictating log\,$R_5(\theta=45\deg) = -2.5$; this yields log\,$F_c =
-3.67$.  In addition we computed the model behaviour for a
$\gamma=4.1$ ($\beta=0.97$) $L=350$~kpc radio source: this model is
shown with the dashed line.

Similarly to Kapahi \& Saikia (\cite{ks82}) and Saikia et al.
(\cite{saiks91}), it is seen that these relativistic jet models --
obviously describing upper limits to linear size distributions --
provide reasonable fits to the data.  It is clear that for the assumed
unbeamed core fraction log\,$F_c = -3.67$ the $\gamma=4.1$ model does
not yield sufficient boosting to explain the high $R_5$ points,
regardless of the inclination of the radio source.  Hence, higher
gamma factors and/or higher unbeamed core fractions must apply.

We conclude that log\,$R_5$ is -- on average -- a good orientation
indicator and that nuclear jet Lorentz factors $\approx$ 4--7 are
likely.  These findings are in broad agreement with results obtained
for samples of (3CR) steep-spectrum quasars and radio galaxies by
Bridle et al. (\cite{bridl94}) and Fernini et al. (\cite{ferni97}),
with the latter authors measuring radio galaxy core fractions
in the range $10^{-2.5}$ -- $10^{-3.5}$.  One caveat should however be
mentioned, namely the possibility of the global radio source
environment affecting the extended radio luminosity, and hence the
$R_5$-parameter.  In addition to core variability, some scatter in $R$
may be attributed to these `local weather' effects (cf. Wills \&
Brotherton \cite{wibroth95}, Barthel \& Arnaud \cite{pdb96}).

\begin{figure*}
 \resizebox{12cm}{!}{\includegraphics{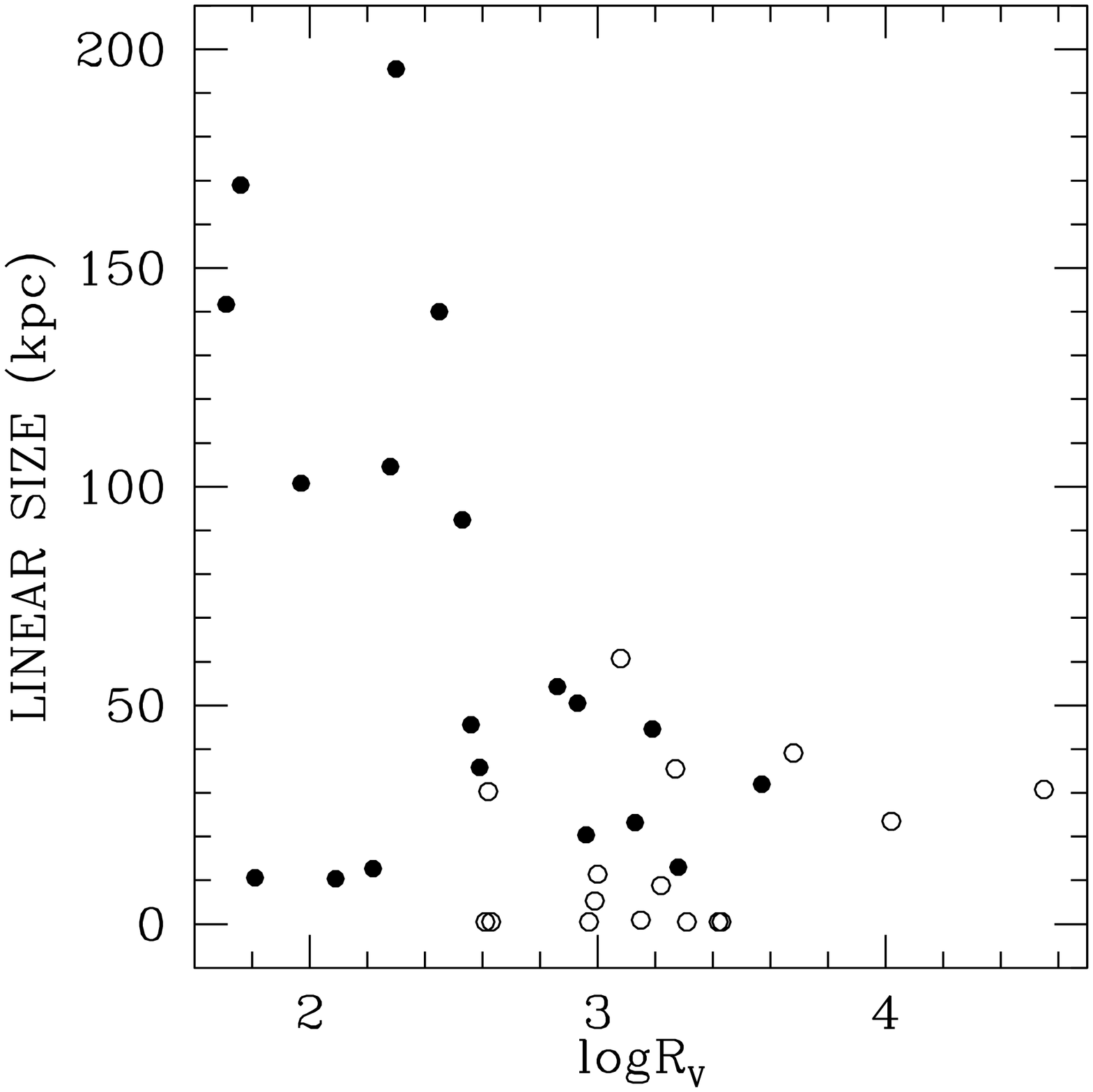}}
 \hfill
 \parbox[b]{55mm}{
  \caption{Linear size vs. V-band luminosity normalized radio core 
   luminosity (5\,GHz) for the combined sample of 35 core- and lobe-dominated 
   quasars from Fig.~\ref{linR5}. See text for details on data points.}
  \label{linRv}}
\end{figure*}

In order to avoid the contamination of a source intrinsic quantity
with an environmental contribution, Wills \& Brotherton
(\cite{wibroth95}) proposed an improved measure of quasar orientation,
$R_V$.  As mentioned already in Sect.~2.4, this $R_V$ parameter is
defined as the ratio of the core radio luminosity at 5~GHz emitted
frequency to the optical V-band luminosity, presumed to be AGN
intrinsic quantities.  Fig.~\ref{linRv} displays the linear size --
log$R_V$ behaviour for the combined sample of 16 flat and 19
steep-spectrum quasars.  It is clear that only compact radio sources
can display strong radio core dominance; large triple quasars display
radio cores of considerably lower (V-band normalised) luminosity.
Fig.~2 represents confirmation of the usefulness of log$R_V$ as quasar
orientation indicator, on the assumption that the upper envelope of
the size distribution measures foreshortening proportional to the
source inclination.

It should be stressed that the above analysis is very simplistic.
Radio sources exist with a range in linear sizes, and a certain spread
in $F_c$ and $\gamma$ must be present, as is evident from Fig.~1.
Nevertheless, this and earlier analyses show that simplistic models
provide agreement with the data, thereby providing support for the
orientation scenario. Obtaining more data points for high $R$ objects
will be most valuable in assessing the magnitude of the beaming
phenomena with more accuracy.

We have obtained optical spectra with the Palomar 200-inch telescope
of 65 sample quasars (Barthel et al. \cite{pdb90}), and have enlarged
this data base at the MMT (Vestergaard \cite{vester99}).  The purpose
of this project is to combine radio and optical data, in a search for
optical orientation indicators.  The log\,$R_5$ and log\,$R_V$
parameters have here once again been shown to be useful orientation
indicators, and we are therefore analyzing the optical spectra to
search for parallel emission line orientation indicators.
Radio-loud quasars offer the advantage that their orientation seems to
be reflected in their Balmer line properties (e.g., Wills \& Browne
\cite{wibrown86}).  We attempt to find similar orientation indicators
in the ultraviolet part of the spectrum, which can be used to address
the orientation of high redshift QSOs in general.  In addition, we
plan to carry out a detailed comparison of quasar emission line
parameters, for radio-loud, radio-quiet, and radio-silent QSOs, in
search for radio-loudness indicators and radio morphological
correlations.  Earlier attempts (e.g., Corbin \& Francis
\cite{corbin94}) suffer from unreliable radio spectral information and
the lack of radio maps. The results of our investigations will be
communicated in forthcoming publications.

\begin{acknowledgements}

We acknowledge the excellent APM facility, CCD imaging by Rien Dijkstra,
and comments on the manuscript by Belinda Wilkes and the referee Dr. 
G.B.~Taylor.  PDB acknowledges travel support from the Leids
Kerkhoven-Bosscha Fonds.  MV is very pleased to thank the Smithsonian
Astrophysical Observatory and the High Energy Division for their
hospitality.  MV also gratefully acknowledges financial support from the
Danish Natural Sciences Research Council (SNF-9300575), the Danish
Research Academy (DFA-S930201) and a Research Assistantship at
Smithsonian Astrophysical Observatory made possible through NASA grants
(NAGW-4266, NAGW-3134, NAG5-4089; P.I.: Belinda Wilkes).  The NRAO VLA
is a facility of the National Science Foundation operated under
cooperative agreement by Associated Universities, Inc.  This research
has made use of the NASA/IPAC Extragalactic database (NED) which is
operated by the Jet Propulsion Laboratory, Caltech, under contract with
the National Aeronautics and Space Administration. 

\end{acknowledgements}

\vfill \eject

\vspace*{4cm}

\appendix

Figs. 3--11 are VLA images at 5~GHz of 9 core-dominated
quasars, with resolving beams shown in each image.  The contour
levels, expressed as multiples of the approximate 1$\sigma$ image
noise level as well as the image peak flux density are specified under
each image.  These numbers are also specified in Table~2. With
reference to Table~1, crosses mark the optical QSO positions with
their associated uncertainties.

\begin{figure*}
 \resizebox{7cm}{!}{\includegraphics{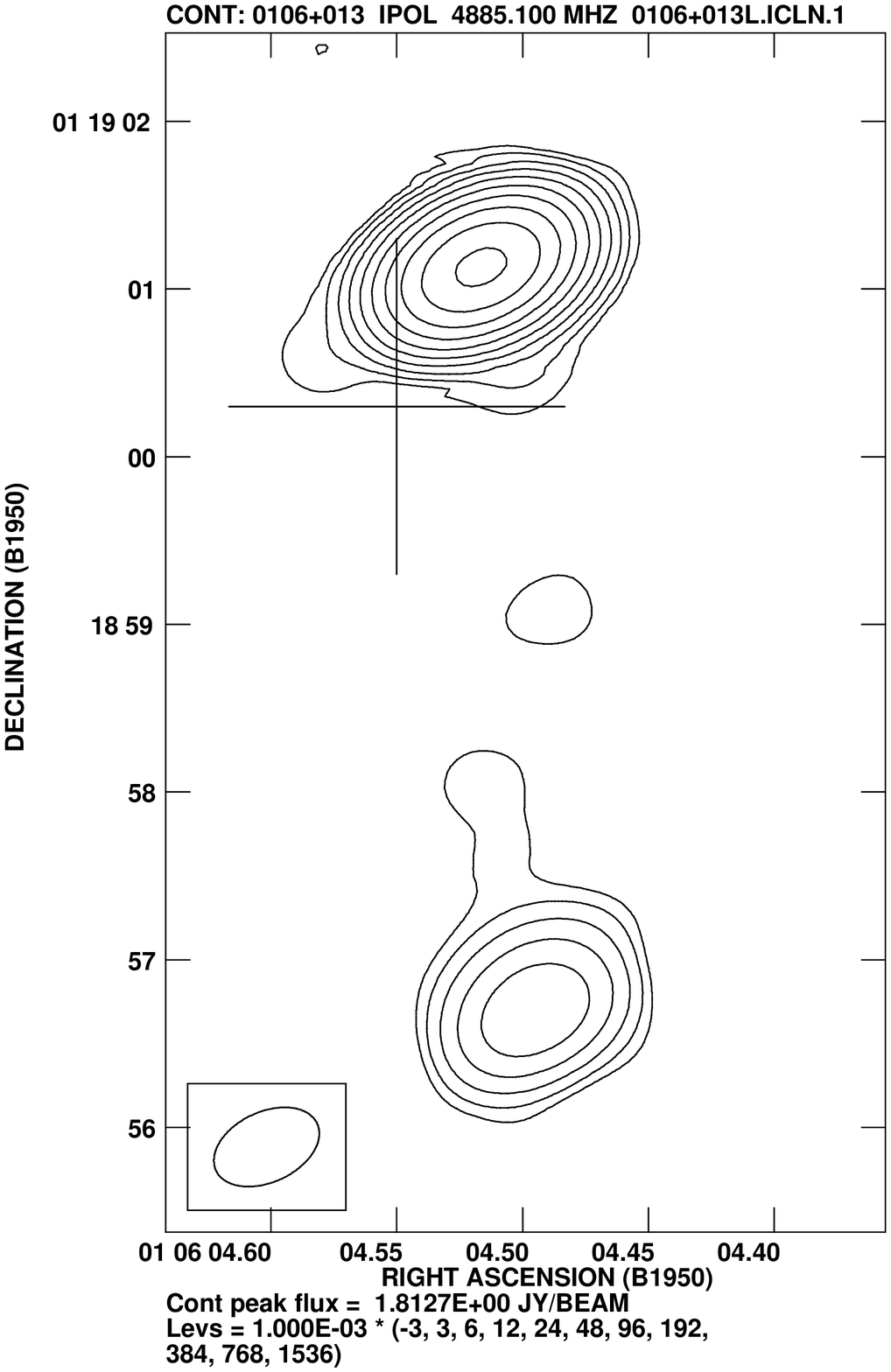}}
 \hfill
 \resizebox{7cm}{!}{\includegraphics{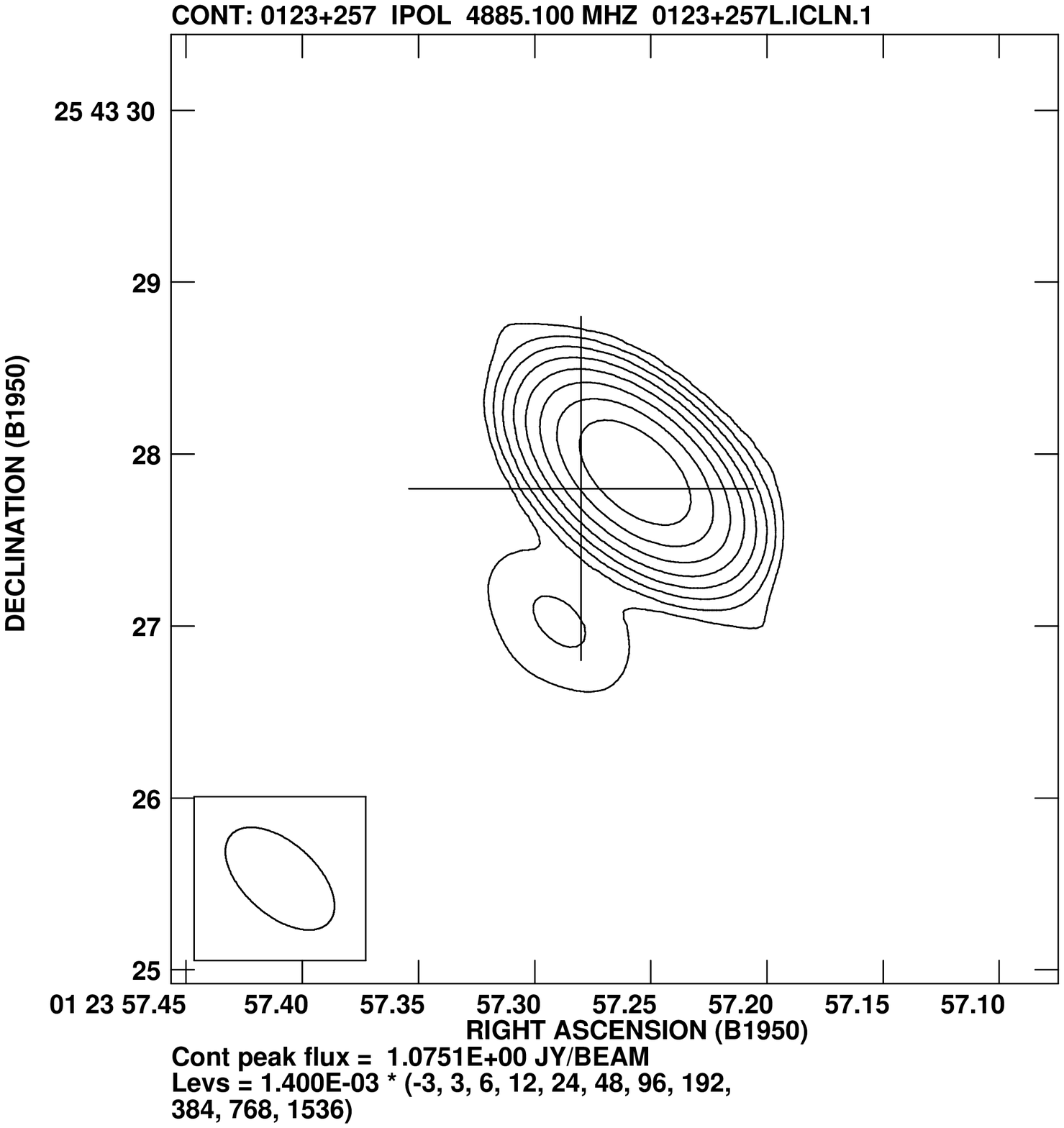}}
\end{figure*}

\begin{figure*}
 \resizebox{7cm}{!}{\includegraphics{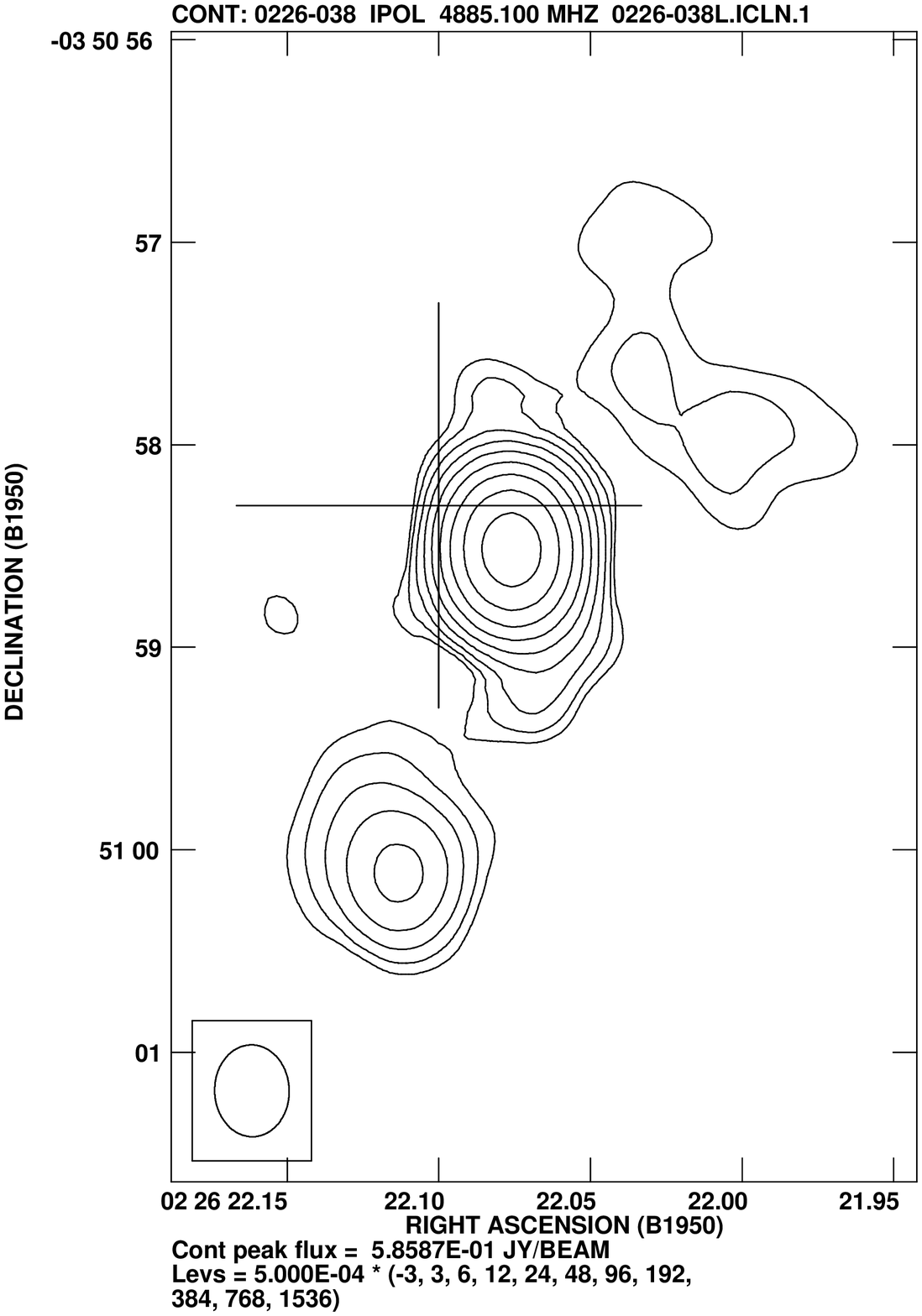}}
 \hfill
 \resizebox{7cm}{!}{\includegraphics{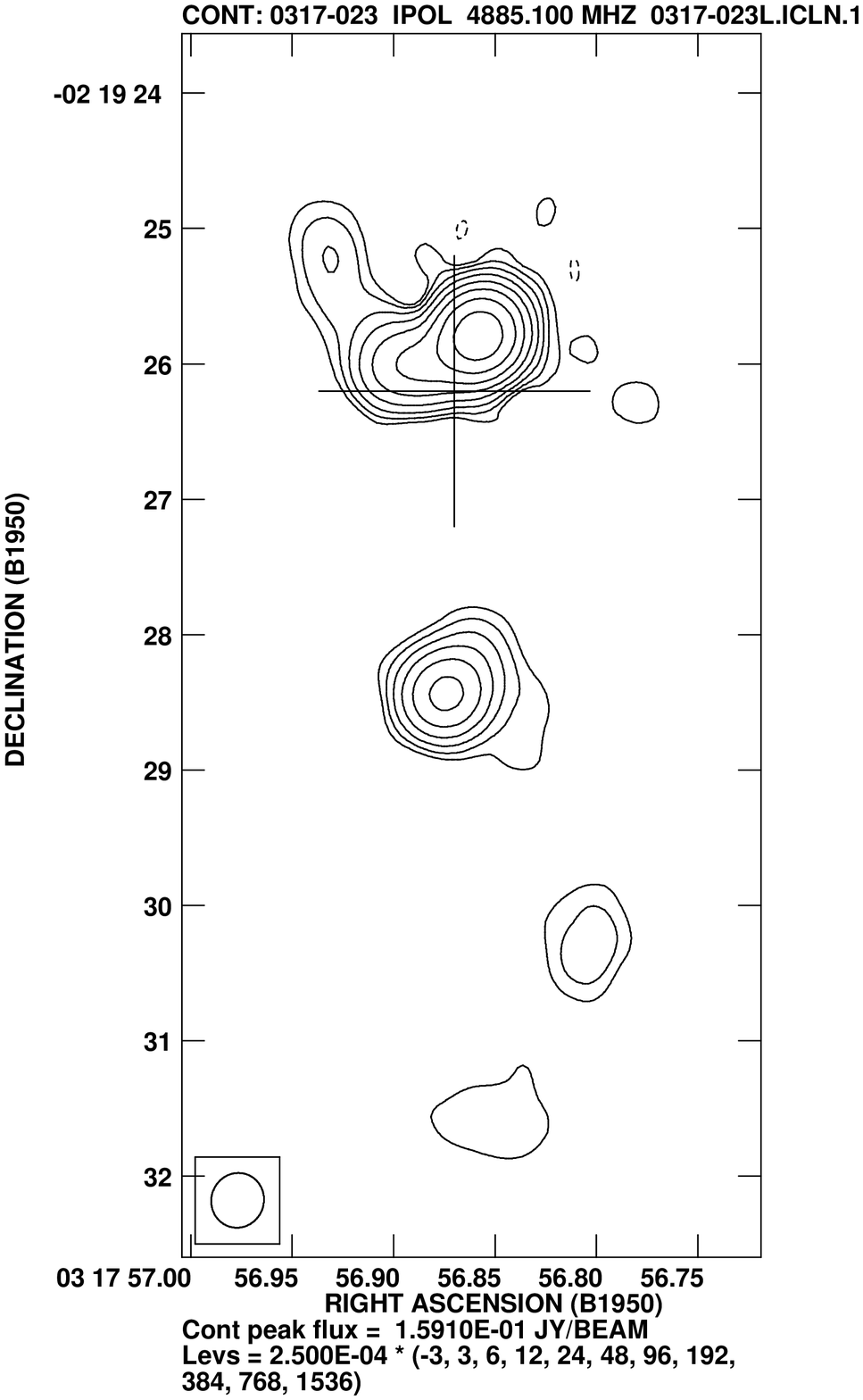}}
\end{figure*}

\begin{figure*}
 \resizebox{7cm}{!}{\includegraphics{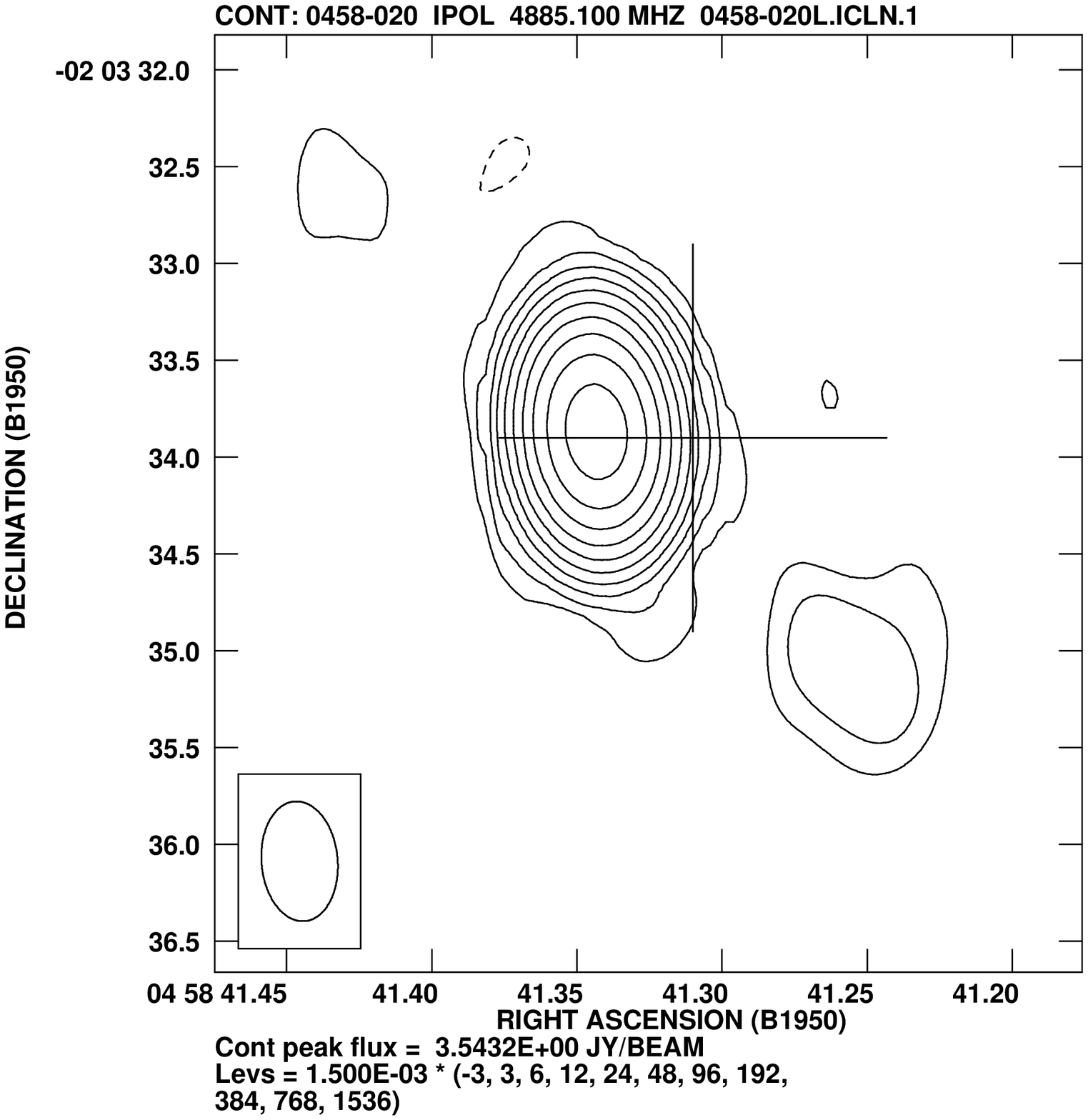}}
 \hfill
 \resizebox{7cm}{!}{\includegraphics{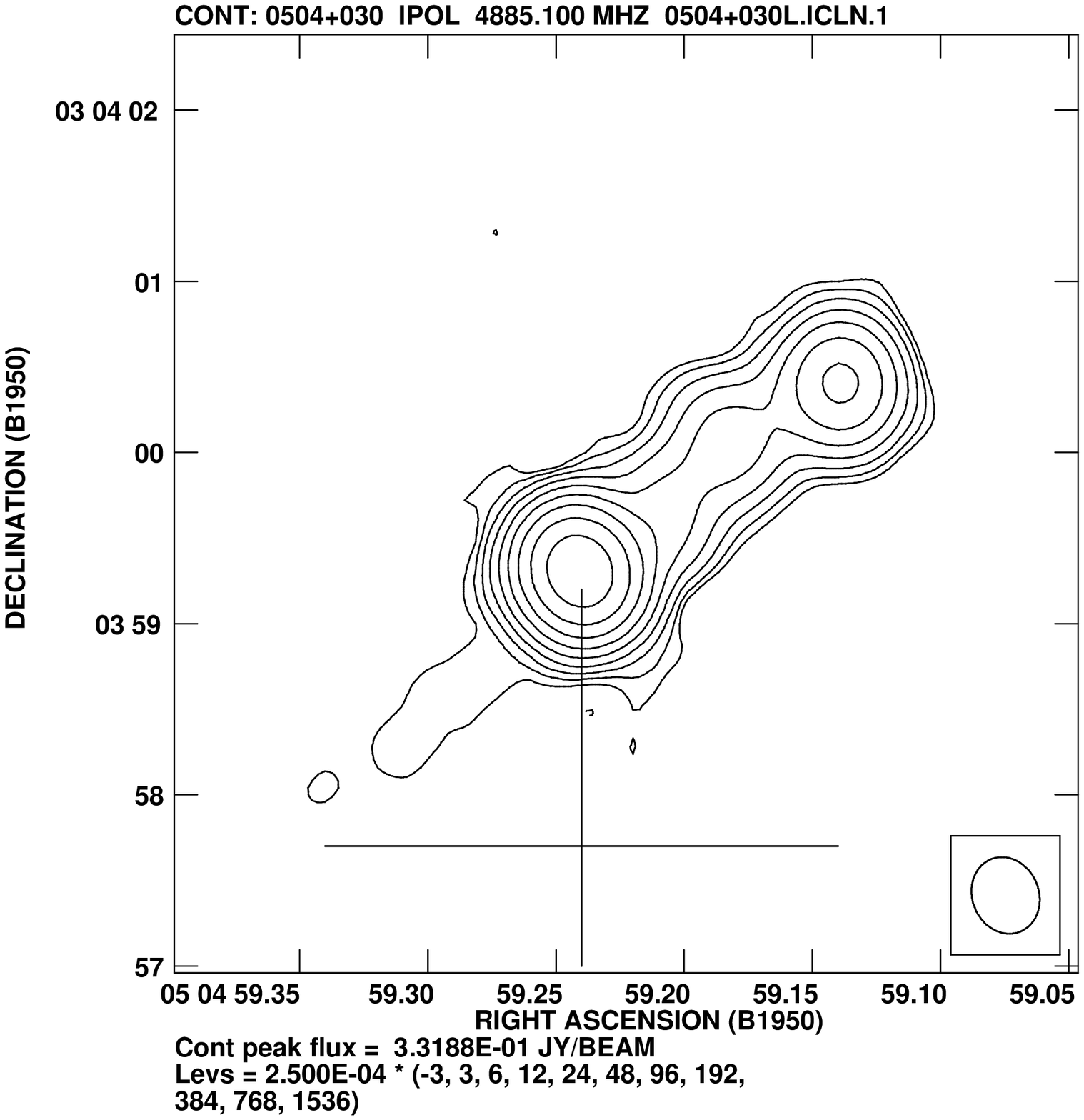}}
\end{figure*}

\begin{figure*}
 \resizebox{7cm}{!}{\includegraphics{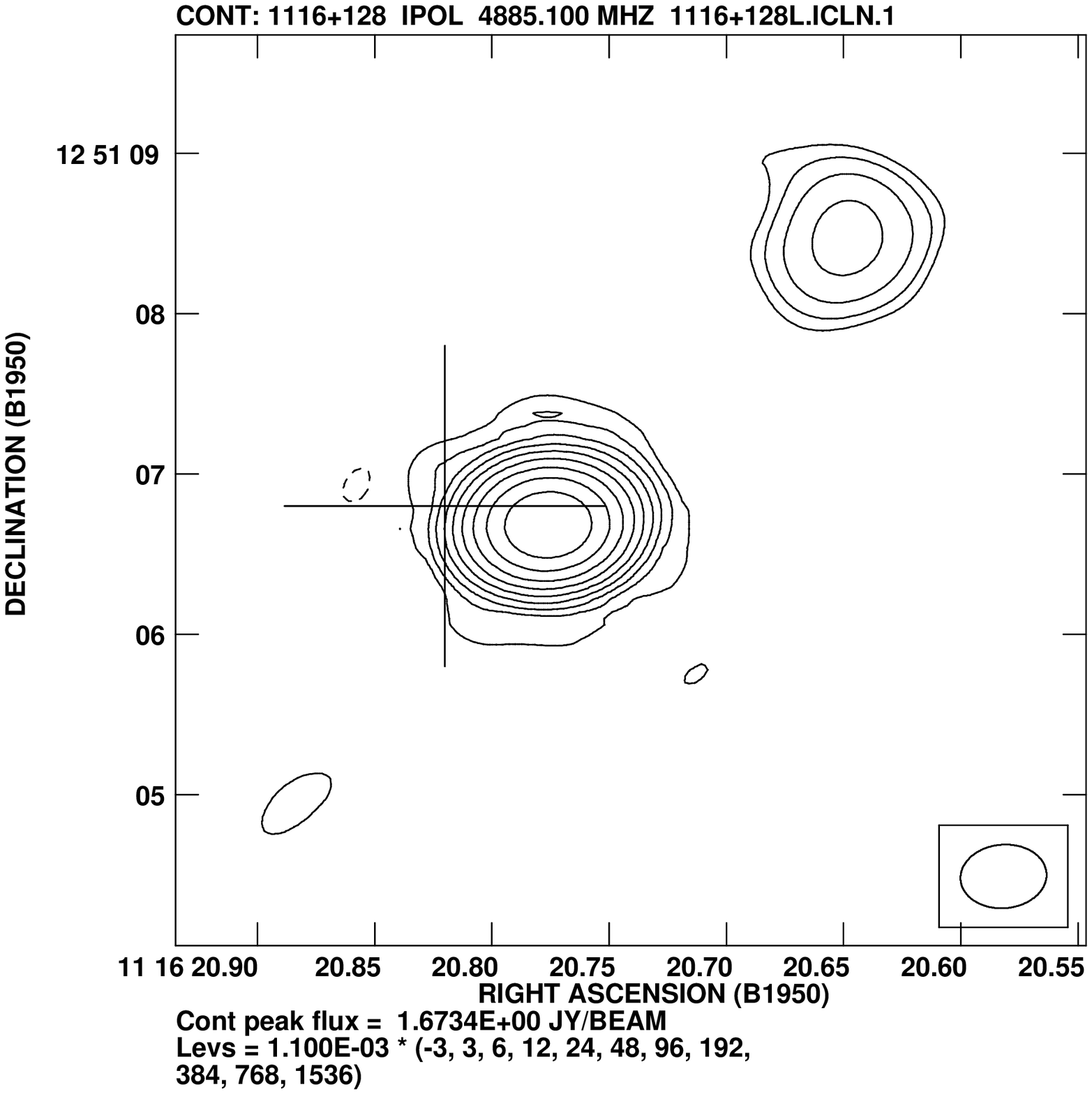}}
 \hfill
 \resizebox{7cm}{!}{\includegraphics{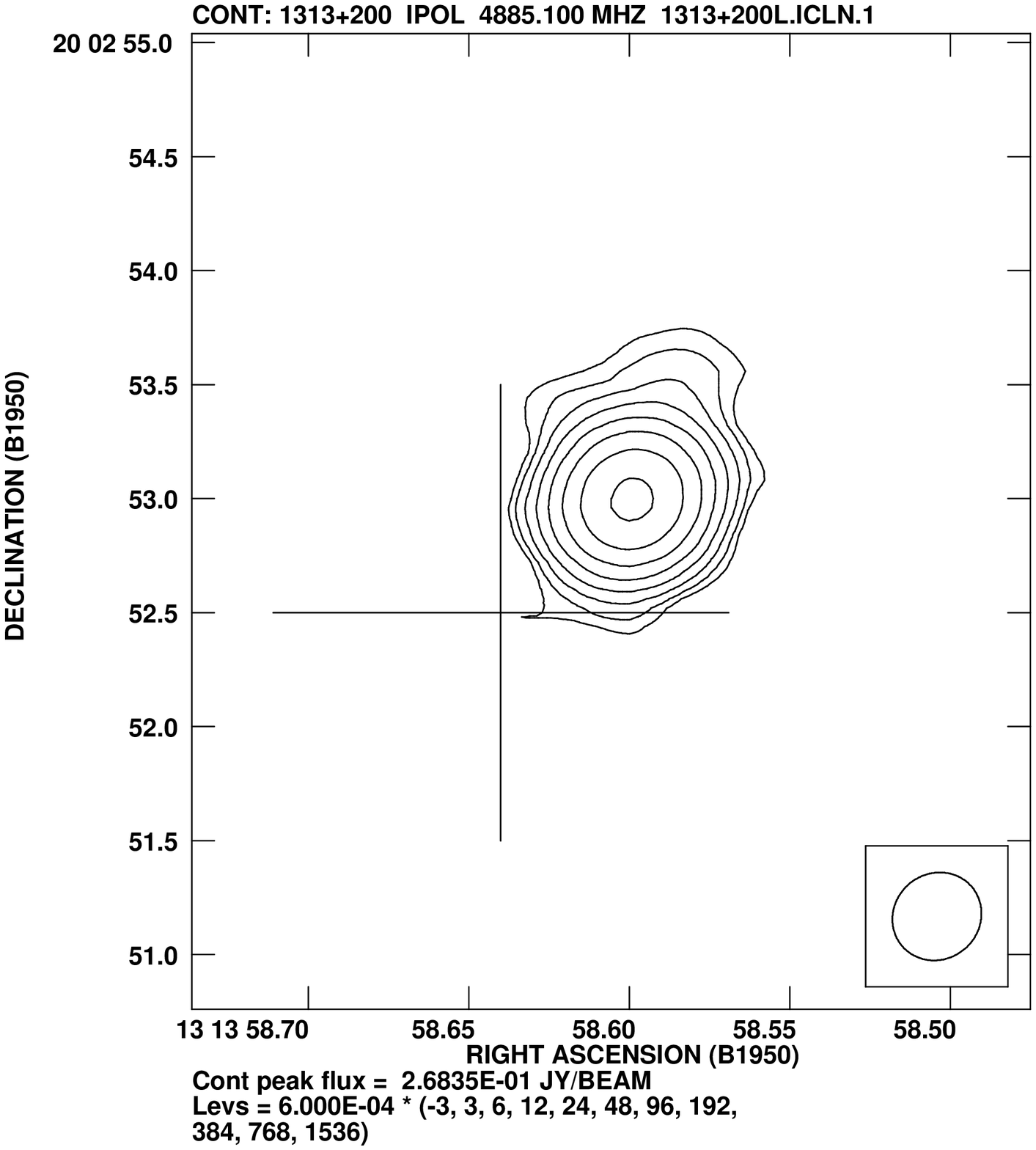}}
\end{figure*}

\begin{figure*}
 \resizebox{7cm}{!}{\includegraphics{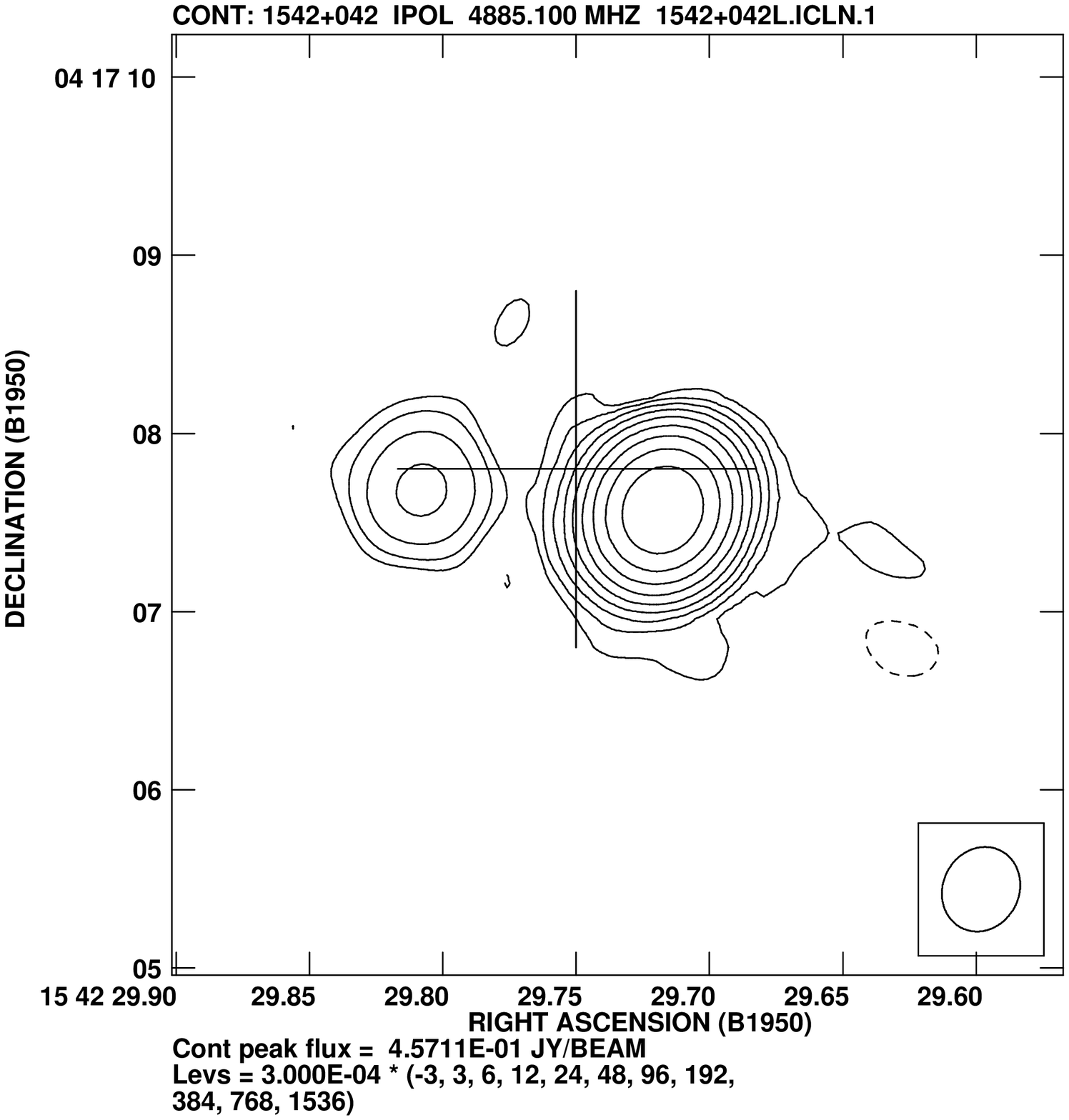}}
\end{figure*}


\end{document}